# A Distributed and Cooperative Approach to Botnet Detection Using Gossip Protocol

**Manoj Rameshchandra Thakur**
Computer Science Department, VJTI, Mumbai, India
manoj.thakur66@gmail.com

**Abstract:** Bots, in recent times, have posed a major threat to enterprise networks. With the distributed nature of the way in which botnets operate, the problems faced by enterprises have become acute. A bot is a program that operates as an agent for a user and runs automated tasks over the internet, at a much higher rate than would be possible for a human alone. A collection of bots in a network, used for malicious purposes, is referred to as a botnet. In this paper we suggested a distributed, co-operative approach towards detecting botnets is a given network which is inspired by the gossip protocol. Each node in a given network runs a standalone agent that computes a suspicion value for that node after regular intervals. Each node in the network exchanges its suspicion values with every other node in the network at regular intervals. The use of gossip protocol ensures that if a node in the network is compromised, all other nodes in the network are informed about it as soon as possible. Each node also ensures that at any instance, by means of the gossip protocol, it maintains the latest suspicion values of all the other nodes in the network.

## 1. Introduction

### 1.1 Related Work:

A number of approaches for intrusion detection have been suggested in the past. Most of these approaches can be categorized into two categories: Artificial Immune System and Soft Computing. A brief overview of these two approaches is as follows:

Approaches based on Artificial Immune System are inspired by the human immune system. [11]. introduces the first lightweight intrusion detection systems based on AIS (Artificial Immune System). [12][13] present intrusion detection system based on 'Danger Theory'. Most of the suggested AIS based Intrusion detection Systems use one of the following algorithms: negative selection algorithm, clonal selection algorithm, artificial immune network, danger theory inspired algorithms and dendritic cell algorithms [14]. [16] focuses on specifically static clonal selection with a negative selection operator. [15] presents a technique to detect anomaly in electromagnetic signals in a complex electromagnetic environment, similar to the complex world wide web.[17] presents an approach to solve the problems with existing intrusion detection systems using autonomous agents. [19] presents a genetic classifier-based intrusion detection system. [18] presents an analogy between the human immune system and the intrusion detection system. It also uses genetic operators like selection, cloning, crossover and mutation attempts to evolve the Primary Immune Responses to a Secondary Immune Response.

The problem of anomaly detection in a system is characterized by lack of exactness and inconsistency. This has encouraged a number of attempts towards intrusion detection based on "Soft

Computing" [20] [21]. "Soft Computing" techniques attempt to evolve inexact and approximate solutions to the computationally-hard task of detecting abnormal patterns corresponding to an intrusion. [22] applies a combination of protocol analysis and pattern matching approach for intrusion detection.[23] proposes an approach towards intrusion detection by analyzing the system activity for similarity with the normal flow of system activities using classification trees. [24] proposes a Soft Computing based approach towards intrusion detection using a fuzzy rule based system. [25] presents an intrusion detection system based on machine learning techniques. [6] presents a proactive detection and prevention technique for intrusions in a Mobile Ad hoc Networks (MANET).[7] suggests a multi-dimensional approach towards intrusion detection which is heuristic in nature and is primarily inspired by earlier works based on soft computing techniques.

Apart from the works mentioned above [10] suggests a distributed technique based on a standalone and network algorithm for botnet detection. This approach is instrumental in detecting and combating bots that operate in a distributed manner.[9] suggests a hybrid approach based on soft computing and artificial immune system. [6] suggests a minor modifications to the existing AODV routing protocol to prevent DDoS attacks in anad hoc wireless network. The suggested approach in [6] shifts the responsibility of monitoring a node of the network to its neighboring node.

The problem of detecting abnormalities and faults in a distributed computing infrastructure is similar to detecting compromised hosts i.e. bots in a given network. Fault tolerant systems successfully employ techniques like state machine replication and gossip protocols to detect failures in their infrastructure [1] [2]. Inspired by the use of such techniques for failure detection and the similarity of the problem with intrusion and abnormality detection, a number of works based on these techniques have been proposed. [8] proposes a novel approach towards intrusion detection based on state machine replication, a commonly used technique for failure detection in fault tolerant systems. In this paper, we propose a technique for botnet detection that exploits the convergent consistency of the gossip protocol to propagate the information related to the abnormality and hence the compromised nature of nodes (bots) in a given network. A detailed explanation of the suggested technique is presented in the following sections of the paper.

**1.2 The Gossip protocol:**

A **gossip protocol** is a computer-to-computer communication protocol inspired by the form of gossip seen in social networks [3]. A number of modern distributed systems employ gossip protocols to solve problems that might be difficult to solve in other ways, primarily because of the distributed nature of the problem. In a gossip protocol, every member of the distributed infrastructure selects a peer at random and shares information specific to the system employing the gossip protocol. For example, in case of fault tolerant systems using the gossip protocol for failure detection, this information contains heartbeat data received by a node from other nodes. The rationale behind using the Gossip protocol is as follows [4] [5]:

1) Gossip protocol provides robust spread of information, any information related to a particular node of a network can be shared across all the nodes of the network very quickly. The time complexity involved in propagating a piece of information among all the node of a network consisting of *n* nodes, assuming that each node gossips with *x* nodes at a time is roughly $O(\log_x n)$.
2) Reliable communication is not a pre-requisite for effective functioning of the gossip protocol.

The approach suggested for detecting botnets in a network is inspired by the GEMS: Gossip-Enabled Monitoring Service for Scalable Heterogeneous Distributed Systems suggested in [2]. The technique used for information exchange between nodes and the procedure used to maintain the updated copy of the information is roughly the same in both the works. However the approach suggested in this paper differs from [2] in the following aspects:

1) The approach suggested in [2] is used for monitoring the health of the system. Whereas the approach suggested in this paper is primarily for botnet detection in a given network.
2) The nodes of the network in [2] exchange heartbeat information among gossip peers.Whereasin the suggested approach, gossip peers exchange suspicion values calculated by the stand-alone agents at each node.

The rest of the paper is structured as follows. Section 2 explains the architecture of the suggested system. Section 3 explains the procedure followed as a part of the gossip protocol in the suggested system followed by the explanation of how confidentiality and integrity is maintained in the suggested system in Section 4. Section 5 summarizes the suggested system followed by the references.

## 2. Architecture

The suggested system to detect botnets in a given network is designed as shown in Figure 1.

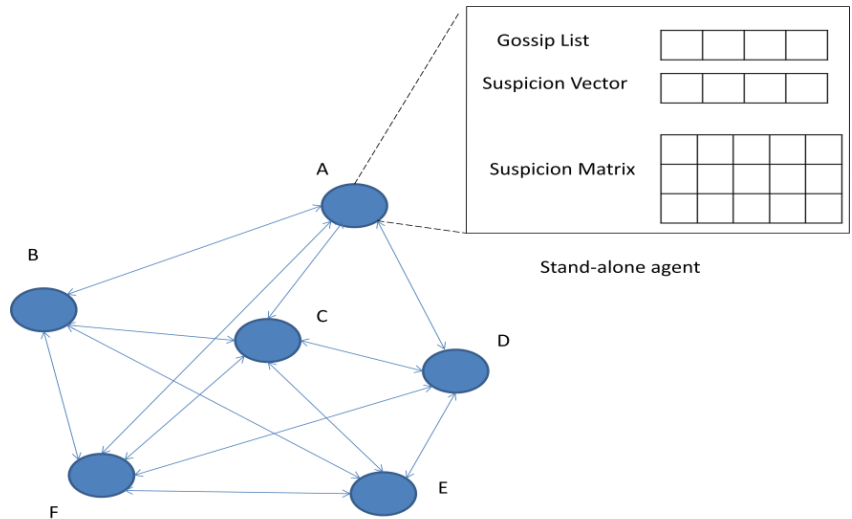

**Figure1 Architecture of the suggested Gossip protocol based system**

Each node in the network runs a *stand-alone agent*. The *stand-alone agent* performs the following functions:
1) Perform analysis of network and system usage statistics of the node on which it runs. This analysis is based on the stand alone algorithm suggested in [10]. The *stand-alone agent* executes the stand alone algorithm after regular intervals and generates a normalized suspicion value (a value between 0 and 5), which it shares with the randomly selected node as a part of the gossip protocol.

The parameters used by the stand –alone agent to infer the suspicion value are as follows:
a. Response time
b. IP addresses
c. Network traffic pattern
d. Ports used for communication
e. Output to input traffic ratio

Stand-alone algorithm suggested in [10] includes a detailed explanation of the above parameters and how the suspicion value is calculated based on these parameters.

2) Maintain and periodically update its own gossip list, suspicion vector and suspicion matrix. The detailed explanation of how this is achievedis presented in Sections 3.

The suggested approach is characterized by the following parameters:

1) $T_{gossip}$ : the time interval between two consecutive gossip messages sent out by a node.
2) $T_{cleanup}$: the interval between the suspicion information was last received for a particular node and the time it is suspected to be compromised.
3) $T_{consensus}$ : time determining how quickly failures are detected and consensus is reached about the compromised nature of a node.

Each node maintains the following data structures to determine whether a given node in the network is compromised or not:

1) Gossip list: a vector containing the number of $T_{gossip}$ intervals since the last gossip packet was received for each node. Thus the lower the value of an index $i$, in the gossip list of a node, the more recently the node had exchanged a gossip packet with the node corresponding to the index $i$.
2) Suspicion vector: a vector constructed such that the $i^{th}$ element of the vector corresponds to the suspicion value inferred by the stand-alone agent of the $i^{th}$ node in the network.
3) Suspicion matrix: an n x n matrix constructed by combining the suspicion vectors from each of the n nodes in the network.

## 3. The Gossip

Each node, after $T_{gossip}$ seconds, selects a peer node from the network at random and exchanges a gossip packet which contains its gossip list and its suspicion vector. The index $i$ in the suspicion vector corresponds to the node itself andcontains the suspicion value inferred by the stand-alone agent of the node itself. Upon reception of a gossip packet, each node updates its suspicion vector based on its own gossip list and the gossip list received from the gossip peer. An illustration of how a node updates its suspicion vector to reflect the latest suspicion value is as follows:

Consider four nodes of a network A, B, C and D. The gossip list, suspicion vector for the nodes A, B and C nodes is follows:

Gossip list [GL]:

| A | NA | 10 | 13 | 10 |
|---|----|----|----|----|
| B | 12 | NA | 15 | 2  |
| C | 20 | 2  | NA | 6  |

Suspicion vector [SV]:

| A | 3.1 | 2   | 2.3 | 2.8 |
|---|-----|-----|-----|-----|
| B | 3.1 | 2.1 | 2.4 | 4.6 |
| C | 3.1 | 2   | 2.3 | 3.7 |

Consider that node A first gossips with node C and receives it's gossip list and suspicion vector as shown above. After $T_{gossip}$ node A gossips with node B and receives it's gossip list and suspicion vector as shown above. Now initially, before these two interactions node A inferred the suspicion value of node D as 2.8. To infer the latest suspicion value of node D, node A first compares the $i^{th}$ entry corresponding to node D in the gossip list of each of the nodes B and C. Since $GL_B[4] < GL_C[4]$ and $GL_A[4] > GL_C[4] > GL_B[4]$ node B has seen the most latest suspicion value for node D, hence node A infers that $SV_B[4]$ is the latest suspicion value for node D.

Thus every node in the network infers its suspicion about a particular node *N* in the network based on the suspicion value received from the node that has most recently interacted with *N*. The advantage of using the gossip approach to propagate the suspicion values is that, if a node is compromised in a network and if one of the nodes in the network detects the compromised nature of that node, the time required to propagate this information throughout the network to all the nodes is considerably lesser. Moreover since the gossip based approach is distributed in nature, the system doesn't suffer from a single point of failure.

### 4. Confidentiality and integrity of the gossip packets:

Confidentiality and integrity of the gossip packets mentioned in section 2 plays an important role in the proper functioning of the suggested system. The use of the gossip protocol ensures eventual consistency for the suspicion value of a node, even in case of an un-reliable communication medium. However if the gossip packets, carrying the gossip list and suspicion vectors, are modified by an unauthorized illegal third party, it would lead to false positives or undetected bot activities.

Confidentiality and integrity of the gossip packets exchanged between the nodes of a network can be ensured by fragmenting and transforming the gossip packets before transmitting them as suggested in [26]. [27] proposes a similar approach towards secure data transfer over a network using the concept of *jigsaw puzzle*. [28] [29] suggest another approach based on the LSB data hiding technique. [30] discusses different steganography techniques that can be applied for ensuring confidentiality and integrity of thegossip packets. In case of the deployment of the suggested system in an ad-hoc mobile network, routing techniques mentioned in [31] can be used for propagating the gossippackets among the nodes of the network.In order to ensure that the gossip packet is exchanged with a legitimate authorized node of the network we use the technique suggested in [32].

## 5. Conclusion

In conclusion the suggested approach is instrumental in providing robust spread of information related to compromised hosts in a given network. The use of the gossip protocol provides the advantage of achieving convergent consistency even in the presence of an un-reliable underlying communication medium. The suggested use of suspicion vectors and gossip list ensures that each of the stand-alone agents work with the latest suspicion values corresponding to the other nodes of the network.